\documentclass[trackchanges]{aastex701}
\usepackage[T1]{fontenc}

\begin{document}

\title{Fast End-to-End Framework for Cosmological Parameter Inference from CMB Data Using Machine Learning}

\author{Larissa Santos}
 \affiliation{Center for Gravitation and Cosmology, Yangzhou University, 
No. 180, Siwangting Road, Yangzhou, 225002, P.R.China}
  \email{larissa@yzu.edu.cn}
\author{Camila P. Novaes}%
\affiliation{%
Instituto Nacional de Pesquisas Espaciais, Av. dos Astronautas 1758, Jardim da Granja, São José dos Campos, SP, Brazil\\
}%
 \email{camila.novaes@inpe.br}
\author{Elisa G. M. Ferreira}
\affiliation{Kavli IPMU (WPI), UTIAS, The University of Tokyo,
5-1-5 Kashiwanoha, Kashiwa, Chiba 277-8583, Japan\\
}%
\affiliation{Center for Data-Driven Discovery, Kavli IPMU (WPI), UTIAS, The University of Tokyo, Kashiwa, Chiba 277-8583, Japan}
\email{elisa.ferreira@ipmu.jp}

\author{Carlo Baccigalupi}
\affiliation{%
  The International School for Advanced Studies (SISSA), via Bonomea 265, I-34136 Trieste, Italy; The National Institute for Nuclear Physics (INFN), via Valerio 2, I-34127, Trieste, Italy; The Institute for Fundamental Physics of the Universe (IFPU), Via Beirut 2, I-34151, Trieste, Italy\\
}
\email{bacci@sissa.it}%


\date{\today}
\begin{abstract}

Precise estimation of cosmological parameters from the cosmic microwave background (CMB) remains a central goal of modern cosmology and a key test of inflationary physics. However, this task is fundamentally limited by strong foreground contamination, primarily from Galactic emissions. In this Letter, we introduce a fast, simulation-based, end-to-end pipeline combining the Analytical Blind Separation (ABS) method for foreground removal with a neural network (NN) framework for cosmological parameter inference. While both tools are individually established, their integration into a single computationally tractable pipeline has not previously been demonstrated. This is primarily enabled by the exceptional computational efficiency of ABS, which, for the first time, makes large-ensemble simulation-based training feasible in the presence of an explicit component separation step. As a proof of principle under controlled full-sky conditions, we assess this framework for the forthcoming LiteBIRD and PICO satellite missions, obtaining $1\sigma$ errors of 0.0035 (LiteBIRD) and 0.0030 (PICO) for the optical depth $\tau$, and 0.0056 (LiteBIRD) and 0.0015 (PICO) for the tensor-to-scalar ratio $r$. Recovered parameters are consistent with input values within $1\sigma$ across most of the test parameter space, with LiteBIRD results consistent with mission forecasts. This demonstrates that the ABS–NN framework provides a computationally efficient and scalable solution for end-to-end CMB inference.

\end{abstract}






\section{\label{intro}Introduction}


The cosmic microwave background (CMB), relic radiation from when the Universe was about 380,000 years old, encodes fundamental information about the early Universe. High-precision measurements of its temperature and polarization anisotropies provide stringent tests of cosmological models and the physics driving cosmic inflation. Upcoming CMB space missions, such as LiteBIRD \citep{LiteBIRD:2020khw}, PICO \citep{NASAPICO:2019thw}, and CORE \citep{Delabrouille:2018}, aim to measure B-mode polarization with unprecedented accuracy, probing the energy scale of inflation. A secondary goal is to achieve cosmic-variance–limited E-mode measurements at low multipoles, improving constraints on the optical depth to reionization, $\tau$, one of the least precisely determined $\Lambda$CDM parameters.

At microwave frequencies, however, the CMB signal is strongly contaminated by foreground emissions from the Galactic interstellar medium and extragalactic sources. Accurate cosmological inference thus depends on effective foreground modeling and removal.  Component separation techniques based on linear combinations of multifrequency data exploit two key properties: (i) distinct frequency dependences of astrophysical emissions and (ii) their statistical independence. Modeling the sky as a mixture of emission components allows component separation  through inversion of a linear system. However, residual contamination in maps and power spectra remains a persistent challenge for future experiments.

Beyond foreground cleaning, precise cosmological parameter estimation and model testing remain central goals of CMB research. Given current tensions within the $\Lambda$CDM framework \citep{DiValentino2025}, alternative inference approaches are both timely and essential. Machine learning (ML), especially neural networks (NNs), has recently emerged as a powerful alternative to traditional likelihood-based methods, enabling simulation-based inference that avoids assumptions about the functional form of the likelihood of the data \citep{2020/cranmer, 2018/alsing, 2019/alsing, 2021/jeffrey}. However, NN-based inference requires extensive training datasets representative of the data distribution, making the repeated application of component separation methods across large ensembles of simulations computationally prohibitive.

A fundamental requirement of simulation-based inference is that the training data represent adequately the actual input to the network at inference time \citep{2020/cranmer, 2018/alsing}. 
Since real CMB observations will undergo component separation before parameter estimation, the training simulations must incorporate the same cleaning step — otherwise a systematic mismatch between training and inference conditions is introduced. However, coupling component separation to large simulation ensembles has remained computationally intractable with conventional map-level techniques. The central novelty of this work is to show that ABS, operating directly at the power-spectrum level with exceptional computational efficiency, removes this barrier and enables, to our knowledge, the first fully integrated pipeline performing component separation and parameter inference in a consistent end-to-end framework at a computational cost tractable for large simulation ensembles.

In this Letter, we propose and validate a fast end-to-end methodology combining the Analytical Blind Separation (ABS) technique ~\citep{Yao:2018,Zhang:2019,Santos:2019}, to estimate cleaned CMB E- and B-mode power spectra, with a NN architecture optimized to infer cosmological parameters, focusing on the optical depth $\tau$ and the tensor-to-scalar ratio $r$ for LiteBIRD- and PICO-like configurations. 
The goal is to demonstrate the feasibility, internal consistency, and statistical performance of the integrated framework under controlled full-sky conditions.


\section{\label{analysis}Analysis}

\subsection{\label{simulation}Simulations}

We generate synthetic polarization maps using the Planck Sky Model \citep[PSM;][]{Delabrouille:2012ye}, which simulates CMB, polarized foregrounds, and instrumental noise (Gaussian  noise, uncorrelated from pixel to pixel and from channel to channel) according to LiteBIRD and PICO specifications (Tables \ref{tab:exp_litebird} and \ref{tab:exp_PICO}). All maps are produced at a HEALPix resolution of $N_{side} = 256$ \citep{Gorski2005}.

We emphasize that the present study adopts white Gaussian noise in a full-sky configuration in order to isolate the performance of the ABS–NN pipeline itself. In realistic satellite observations, large-scale modes are affected by $1/f$ noise and by scan-strategy–induced effects. The incorporation of these effects requires a consistent treatment of time-ordered data, map-making systematics, and partial-sky masking, which in turn introduces E–B leakage and mode-coupling effects. Addressing these in a self-consistent way necessitates additional formalism and it is reserved for future work.

\subsubsection{\label{simulation_maps}CMB maps}

We compute theoretical $EE$ and $BB$ angular power spectra using the Boltzmann solver CLASS \citep{2011/lesgourgues, 2011/blas} under the $\Lambda$CDM framework. To train and test the NN for cosmological inference, we calculate $C_\ell^{EE}$,$C_\ell^{BB}$ for a set of 700 different \{$r$, $\tau$\} pairs of values (each pair refereed as a cosmology). The training set spans $0 < r < 0.05$ and $0.01 < \tau < 0.13$  (500 samples), while the test set covers $0 < r < 0.03$ and $0.038 < \tau < 0.070$  (200 samples).
This strategy ensures that the NNs are trained over sufficiently broad parameter ranges, mitigating the prior edge effects that could otherwise introduce biases in the inferred parameters \citep{2022/villaescusa, 2022/perez}. The narrower parameter range adopted for the test set, while avoiding such boundary-induced artifacts, confines the inference to physically plausible regions of the parameter space consistent with current cosmological constraints \citep{2020/planck-vi}.

\begin{figure*}
\includegraphics[width=\textwidth]{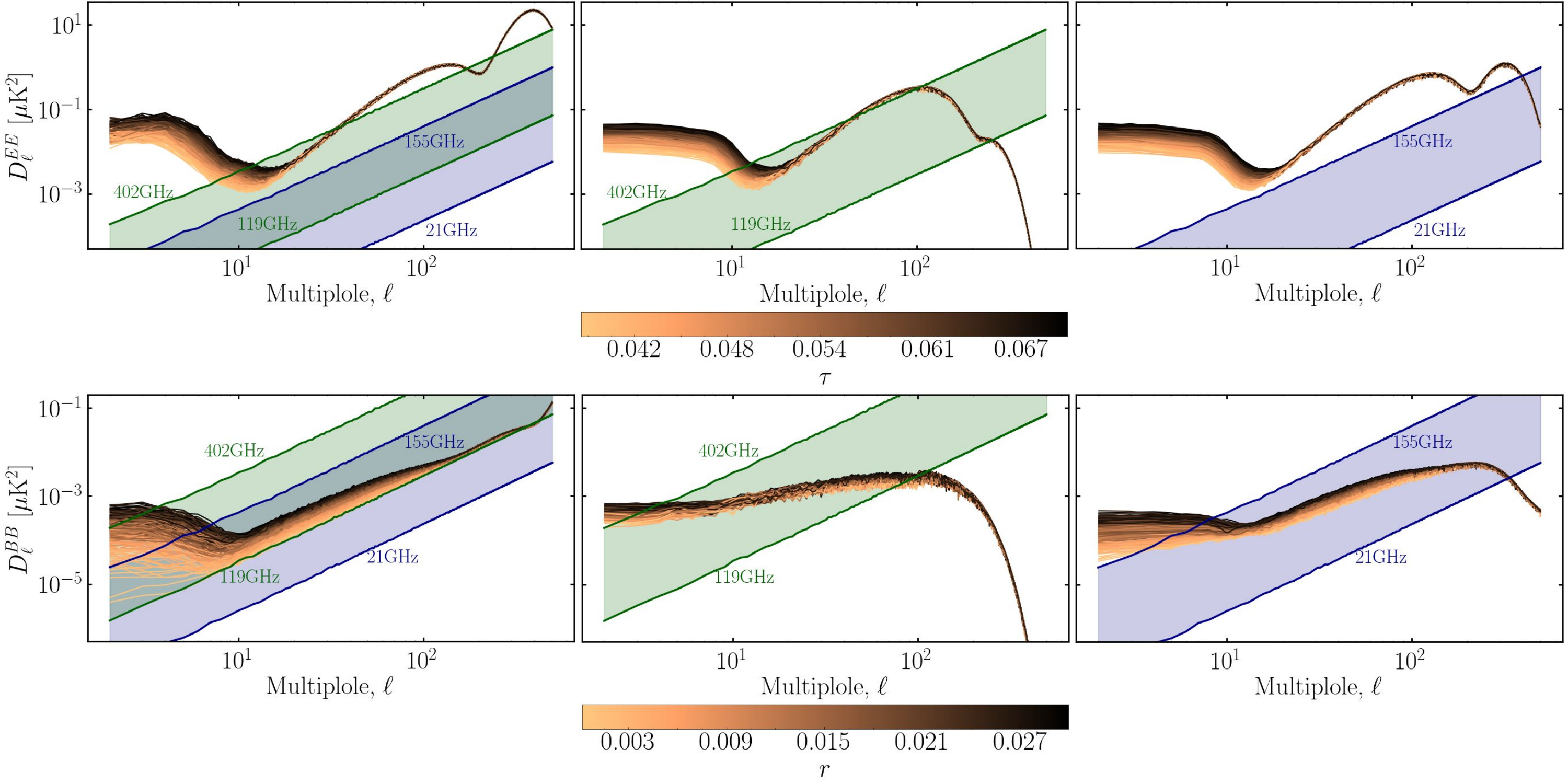}
\caption{\label{fig:Cls} Angular power spectra of the input clean CMB simulations (left) and the spectra recovered by ABS from LiteBIRD- (middle) and PICO-like (right) simulations. ABS spectra include the effect of the corresponding largest beam (first row of Tables \ref{tab:exp_litebird} and \ref{tab:exp_PICO}).  The green and blue bands represent the range of noise levels across frequency channels for LiteBIRD and PICO, respectively, bounded by the channels with minimum and maximum noise levels.
}
\end{figure*}

We sample the parameter space using the Latin Hypercube approach \citep{1979/mckay_LH} and generate 10 realizations of Q and U polarization maps for each cosmology using the power spectra. All other $\Lambda$CDM parameters are fixed to the Planck 2018 best-fit values: $H_0 = 67.27$ km s$^{-1}$, $\Omega_c = 0.265$, $\Omega_b = 0.049$, $n_s = 0.9649$. Given the constraint $10^9 A_s e^{-2\tau} = 1.884$ \citep{wolz}, we fix this combination while varying $A_s$ and $\tau$. The resulting $E$- and $B$-mode spectra (Figure \ref{fig:Cls}, left) illustrate the impact of varying $r$ and  $\tau$.

\subsubsection{\label{simulation_foreground}Foreground contaminants}

In addition to the CMB, several astrophysical processes contribute to sky emission across 21–462 GHz. We model the two dominant sources of polarized Galactic foregrounds: synchrotron radiation from relativistic  charged particles in Galactic magnetic fields, dominant at low frequencies, and thermal dust emission from heated dust grains, prevailing at high frequencies. Extragalactic contributions from radio and far-infrared galaxies, along with cluster emission, are also included to ensure a comprehensive description of polarized contaminants.


\begin{table}[htpb]
\begin{minipage}[b]{90mm}
\small
\caption{LiteBIRD specifications \citep{LiteBIRD:2020khw, LiteBIRD:2022cnt}.}
\label{tab:exp_litebird}
\begin{tabular}{ccc}
\hline \hline
Band center & Beam FWHM&  noise level\\
(GHz)     &   (arcmin)       & ($\mu$K${}_{\rm CMB}$-arcmin) \\
\hline
040 &  70.5  & 37.42 \\
050   & 58.5  & 33.46 \\
060   & 51.1   & 21.31  \\
068   & 47.1  & 16.87  \\
078   & 43.8  & 12.07  \\
089  &  41.5  & 11.3  \\
100  &  37.8  & 6.56 \\
119  &  33.6  & 4.58  \\
140  &  30.8  & 4.79  \\
166 &  28.9  & 5.57 \\
195   & 28.6  & 5.85 \\
235   & 24.7   & 10.79  \\
280   & 22.5  & 13.80  \\
337   & 20.9  & 21.95  \\
402  &  17.9  & 47.45  \\
\hline
\end{tabular}
\end{minipage}
\begin{minipage}[b]{90mm}
\caption{PICO specifications \citep{NASAPICO:2019thw}.}
\label{tab:exp_PICO}
\begin{tabular}{ccc}

\hline \hline
Band center & Beam FWHM&  noise level\\
(GHz)     &   (arcmin)       & ($\mu$K${}_{\rm CMB}$-arcmin) \\
\hline
021 &  38.4  & 16.9 \\
025   & 32.0  & 13.0 \\
030   & 28.3   & 8.7  \\
036   & 23.6  & 5.6  \\
043   & 22.2  & 5.6  \\
052  &  18.4  & 4.0  \\
062  &  12.8  & 3.8 \\
075  &  10.7  & 3.0  \\
090  &  9.5  & 2.0  \\
108 &  7.9  & 1.6 \\
129   & 7.4  & 1.5 \\
155   & 6.2   & 1.3  \\
186   & 4.3  & 2.8  \\
223   & 3.6  & 3.2  \\
268  &  3.2  & 2.2  \\
321  &  2.6  & 3.0  \\
385  &  2.5  & 3.2  \\
462  &  2.1  & 6.4  \\

\hline
\end{tabular}
\end{minipage}
\end{table}

\subsection{\label{comp_sep}Component separation}

A novel approach \citep{Zhang:2019} enables direct estimation of the CMB angular power spectrum from multivariate spectra of multifrequency data. Its computational simplicity and efficiency make it ideal for integration with neural network–based parameter inference in an end-to-end framework. The ABS method has been validated on simulated datasets \citep{Yao:2018,Santos:2019} and implemented in the AliCPT-1 pipeline \citep{Zhang:2024}, a high-altitude CMB polarization experiment, demonstrating its robustness and readiness for future satellite missions.

The ABS formalism assumes a data model in which the observations in $N_f$ different frequency channels contain a superposition of CMB, foreground emission, and noise as 
\begin{equation}
\label{eq:data}
\mathcal{D}^{\rm obs}_{ij}(\ell) = f_if_j\mathcal{D}^{\rm cmb}(\ell) + \mathcal{D}^{\rm fore}_{ij}(\ell) + \delta \mathcal{D}_{ij}^{\rm noise}(\ell)\,,
\end{equation}
where $i,j = 1,2, ..., N_f$.  We use thermodynamic units for the observations, so that the
CMB emission pattern is constant across frequencies, and $f_i=1$,  $\forall i$. $\mathcal{D}^{\rm obs}_{ij}$ represents the cross-band power spectrum of the observations in the $i$- and $j$-th frequency channels. The three main contributions to the data are the CMB signal $\mathcal{D}^{\rm cmb}$, $\mathcal{D}^{\rm fore}_{ij}$ and $\delta \mathcal{D}_{ij}^{\rm noise}$, which are the cross band power matrices of the foreground and residual instrumental noise, respectively. The ensemble-averaged noise power spectrum is assumed to be known and has been subtracted beforehand from the measured cross-power spectrum. The ABS solution can be written as:
\begin{equation}
\label{eq:abs2}
\widehat{\mathcal{D}}^{\rm cmb} = \left( \sum^{{\tilde{\lambda}}_{\mu}\geq \lambda_{\rm cut}} \tilde{G}^2_{\mu}\tilde{\lambda}_{\mu}^{-1}\right)^{-1}\ - \mathcal{S},
\end{equation}
where
\begin{equation}
\label{eq:noiseD}
\mathcal{\widetilde{D}}^{\rm obs}_{ij}\equiv \frac{\mathcal{D}^{\rm obs}_{ij}}{\sqrt{\sigma_{\mathcal{D},i}^{\rm noise}\sigma_{\mathcal{D},j}^{\rm noise}}} + \tilde{f}_i\tilde{f}_j\mathcal{S}\,\\
\end{equation}

\begin{equation}
\tilde{f_i} \equiv \frac{f_i}{\sqrt{\sigma_{\mathcal{D},i}^{\rm noise}}}\,,~\tilde{G}_{\mu}\equiv {\bf \tilde{f}}\cdot {\bf \tilde{E}}^\mu\,.
\end{equation}
Here ${\bf \tilde{E}}^\mu$ and $\tilde{\lambda}_\mu$ are the $\mu$-th eigenvector and corresponding eigenvalue of $\mathcal{\widetilde{D}}^{\rm obs}_{ij}$. The ABS method thresholds the eigenvalues $\tilde{\lambda}_\mu$, only keeping those signal-dominated modes. We choose $\tilde{\lambda}_{\rm cut} = 1$ for $EE$ and $BB$ spectra according to ~\cite{Yao:2018}. In Eq.~\ref{eq:abs2}, the free parameter $\mathcal{S}$ corresponds to an amplitude shift of the input CMB power spectrum from $\mathcal{D}^{\rm cmb}$ to $\mathcal{D}^{\rm cmb} +\mathcal{S}$,  particularly important for low signal-to-noise regime.

The rapid implementation of ABS is the critical enabling factor that makes this end-to-end ML training practical. Previously, incorporating full component separation into a large simulation-based ML pipeline would have required prohibitive computational resources.

\subsection{\label{parameters}Cosmological constraints}

We use NNs to map the EE and BB power spectra recovered by ABS in a full-sky analysis for LiteBIRD and PICO (second and third columns of Figure \ref{fig:Cls}) into the parameters of interest, $r$ and $\tau$. 
The cosmological inference follows a procedure fully based on simulations, as described below \cite[for further details, see][]{2024/novaes, 2024/novaes-hsc}.

\subsubsection{Training and test data sets}

The statistical performance of simulation-based inference is directly tied to both the amount of synthetic data available and how well the training data reproduce the conditions under which the network is applied.
We address these requirements through the ABS method, which operates directly at the level of the angular power spectrum rather than on sky maps,  greatly improving computational efficiency over conventional component-separation techniques. This enables the rapid generation of the extensive and statistically representative training and test data sets required for robust  ML–based cosmological inference.
The goal here is not to test robustness against unknown systematics, but to evaluate the 
internal consistency and statistical performance of the integrated ABS–NN pipeline under controlled conditions. A detailed analysis of robustness under model mismatch (e.g., spatially varying foregrounds, correlated noise, or imperfect noise subtraction within ABS) will be presented in follow-up work.

Here, we use a set of 10 recovered ${BB, EE}$ spectra for each of the 500+200 cosmologies, or pairs \{$r, \tau$\}, considered for simulations.
We split the first 500 into 400/100 cosmologies training/validation of the NN model, testing it on the other 200 cosmologies (narrower intervals of parameters). 
This ensures that cosmologies used for testing are entirely independent from those employed in training and validation. A cross-validation procedure is implemented following \cite{2024/novaes}. 


The training+validation set, $\mathcal{T}(X^i,y^i)$, is defined so that the NN can learn the relation between the BB (EE) power spectrum, $X^i = C_\ell^{BB,i}$ ($X^i = C_\ell^{EE,i}$) and $r$ ($\tau$) parameter, $y^i = r^i$ ($y^i = \tau^i$), for the $i$th simulation.
Notice that the training is performed over (i) $\mathcal{T}(C_\ell^{BB,i},r^i)$ and (ii) $\mathcal{T}(C_\ell^{EE,i},\tau^i)$ separately, resulting in an NN model for each case.
Since this procedure is applied independently for LiteBIRD and PICO, this results in four NN models in total: one per combination of instrument and power spectrum.
The efficiency of the trained NN models is evaluated over the test set, $t(X^{j})$, outputting the predicted values of the corresponding cosmological parameter $y^j$ for the $j$th simulation of the test set. 

\subsubsection{Neural network implementation}

We employ a fully connected NN, using the \textsc{Optuna} package \citep{2019/akiba-optuna} to sample the hyperparameters space and automatically define the optimal architecture \footnote{We fix the optimizer to {\tt Adam}, with $\beta_1,\beta_2 = 0.9, 0.999$. The hyperparameters and the respective searching spaces considered by \textsc{Optuna} are: number of neurons in the hidden layers: $[1,500]$; number of layers: $[1,3]$; activation function: [{\tt ReLu}, {\tt tanh}]; learning rate: $[10^{-4},10^{-2}]$; batch size: $[50,500]$; and number of epochs: $[50,500]$. }. 
The maximal number of trials (tested architectures) is fixed to 500, regardless of the case, (i) or (ii), which usually takes no more than 24 hours on 56 cores of a processor Intel Xeon Gold 5120 2.20 GHz and 512 GB of RAM. 
After defining the architecture, the training+validation process provides an NN model in a few minutes. 
The loss function is chosen to be the mean square error ({\sc mse}), $\mathcal{L} = \langle (y^{Pred} - y^{True})^2 \rangle$, averaged over the simulations, where $y^{Pred}$ and $y^{True}$ are the predicted and true values of the cosmological parameter to be constrained. 
The final hyperparameters selected by \textsc{Optuna} for each of the four NN models are reported in Table \ref{tab:hyperparams}. 
No preprocessing of the input power spectra was applied prior to training; the spectra are passed directly to the network without normalization, PCA decomposition, or dimensionality reduction. The stability of the optimization is confirmed by the close agreement between training and test {\sc rmse} values reported in Table \ref{tab:results} (see Section \ref{sec:inference}).

\begin{table}[htpb]
\caption{Final hyperparameters selected by \textsc{Optuna} for each of the 
four NN models. For networks with two layers ($EE$ networks), the number of neurons reported corresponds to the two separate layers. \label{tab:hyperparams}}
\begin{tabular}{lcccccc}
\hline
\hline
Model & Layers & Neurons & Activation & Learning rate & Batch size & Epochs \\
\hline
LiteBIRD-BB ($r$)   & 1 & 2 & {\tt tanh} & 1.38$\times 10^{-3}$ & 133 & 341 \\
LiteBIRD-EE ($\tau$) & 2 & 43/361 & {\tt tanh} & 7.45$\times 10^{-4}$ & 69 & 421 \\
PICO-BB ($r$)        & 1 & 4 & {\tt ReLu} & 2.92$\times 10^{-4}$ & 66 & 483 \\
PICO-EE ($\tau$)     & 2 & 11/249 & {\tt ReLu} & 1.47$\times 10^{-3}$ & 86 & 485 \\
\hline
\end{tabular}
\end{table}

\begin{figure}[t]
\small
\includegraphics[width=0.45\columnwidth]{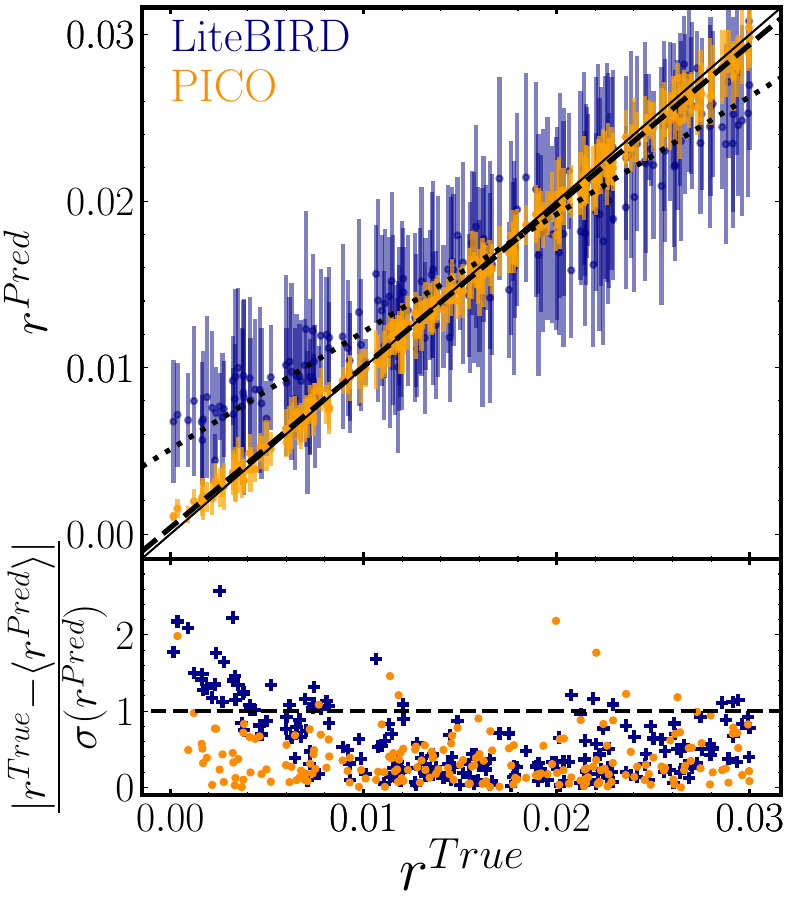}
\includegraphics[width=0.45\columnwidth]{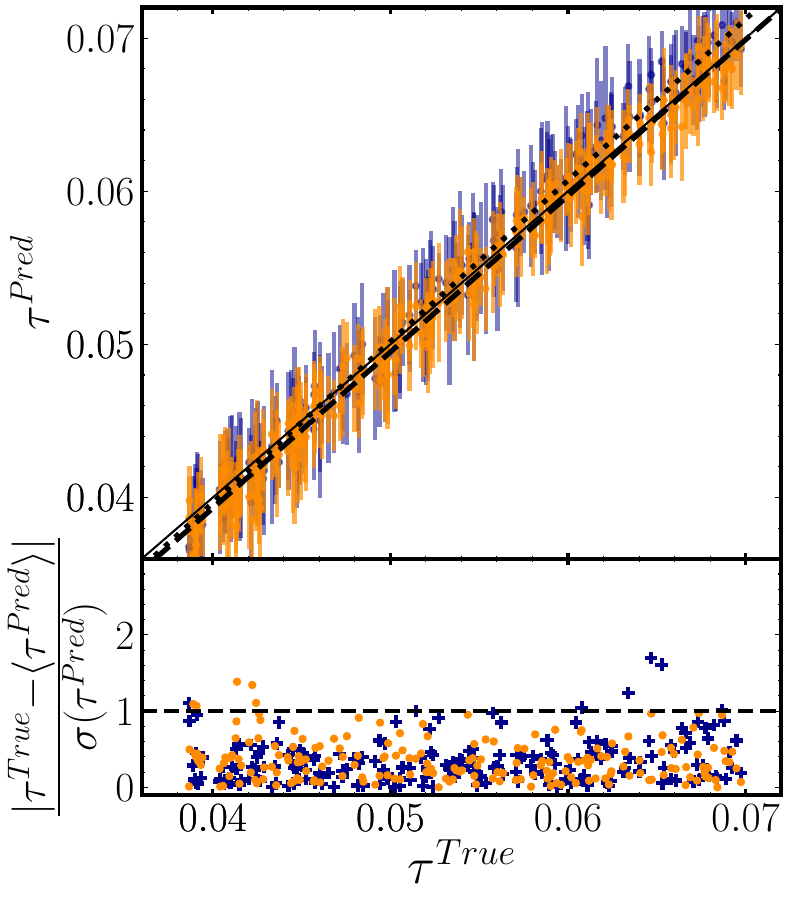}
\caption{\label{fig:predXtrue} Predicted versus true cosmological parameters ($r$ and $\tau$) for LiteBIRD-like (blue) and PICO-like (orange) instruments. Dots and error bars show the mean and standard deviation from 10 simulations for each of the 200 cosmologies of the test set. The black dotted and dashed lines denote the linear fits to LiteBIRD and PICO results, while the thin diagonal indicates the identity $y^{Pred}=y^{True}$. The corresponding root-mean-square errors are also shown. Bottom panels display the statistical significance of the predictions as a function of the true values, with the dashed line marking the 1$\sigma$ level.}
\end{figure}

\subsubsection{Parameter inference} \label{sec:inference}

Figure \ref{fig:predXtrue} presents the results of applying the trained NN models to the test sets, comparing the predicted and  true values of $r$ and $\tau$.
Model performance is quantified using two error estimates: (i) the average standard deviation over the 200 cosmologies of the test set, $\langle \sigma \rangle = \sigma(y^{Pred})$, where $\sigma$ is the standard deviation over the 10 predictions for one cosmology (ii) the root-mean-square error comparing predicted and true values of the 10 $\times$ 200 simulations, {\sc rmse} $= \sqrt{\mathcal{L}}$, corresponding to the $1\sigma$ error from all predictions. 
Table \ref{tab:results} summarizes these results. 
The close agreement between the {\sc rmse} calculated from the training and test sets confirms that our results are not affected by overfitting. 
We note that these uncertainty estimates are derived entirely within the controlled simulation framework and should be interpreted as lower bounds on the uncertainties expected when applied to real data. 
More principled uncertainty quantification can be obtained through approaches such as Bayesian neural networks or normalizing flows, which is reserved for follow-up work.

\begin{table}[t]
\caption{\label{tab:results}
Error estimates evaluating the performance of our predictions of the cosmological parameters. In parenthesis we show \textsc{rmse} estimated from the training set.
}
\begin{ruledtabular}
\begin{tabular}{ccccc}
 & \multicolumn{2}{c}{\textrm{LiteBIRD}} &  \multicolumn{2}{c}{\textrm{PICO}} \\
 & $\langle \sigma \rangle \times 10^{2}$ & \textsc{rmse} $\times 10^{2}$ & $\langle \sigma \rangle \times 10^{2}$ & \textsc{rmse} $\times 10^{2}$ \\
\colrule
 $r$    & 0.50 & 0.56 (0.52) & 0.10 & 0.15 (0.15) \\
 $\tau$ & 0.30 & 0.35 (0.33) & 0.30 & 0.30 (0.27) \\
%
\colrule 
\end{tabular}
\end{ruledtabular}
\end{table}

\section{Discussion and conclusion} \label{sec:discussion}


In this work, we introduce and validate a novel integrated ABS–NN framework for cosmological parameter inference. While both ABS and ML-based inference are established tools individually, their combination into a single computationally viable end-to-end pipeline constitutes a new practical advance. This integration enables fast, large-ensemble simulation-based inference that includes explicit foreground separation, which has not been previously realized due to computational limitations. The exceptional efficiency of ABS, capable of cleaning foregrounds from hundreds of simulated skies within a few days, makes it ideally suited for simulation-based inference in CMB analyses.

As shown in Figure \ref{fig:Cls}, ABS robustly recovers both $EE$ and $BB$ power spectra, with PICO-like instruments slightly outperforming LiteBIRD-like configurations in B-mode reconstruction, consistent with their noise levels. The resulting parameter constraints (Figure \ref{fig:predXtrue}) confirm this performance: $r$ is recovered with $1\sigma$  uncertainties of $0.0056$ for LiteBIRD and $0.0015$ for PICO, fully consistent with mission specifications. 
For PICO, $r$ values are recovered within $1\sigma$ across the full test parameter range, while for LiteBIRD this holds in the regime $r > 0.01$, with larger deviations at $r < 0.01$ where the signal-to-noise ratio is lowest. 
Note that the systematic deviation in the LiteBIRD BB predictions at the lowest and highest $r$ is a known effect of {\sc mse}-based regression in low signal-to-noise regimes, where the network minimizes the loss by predicting values close to the mean of the training distribution  \citep{2022/perez, 2024/novaes, 2024/novaes-hsc}, and is not specific to the ABS-reconstructed spectra.

 
From the $EE$ spectra, both missions yield consistent $\tau$ constraints with $1\sigma$ errors of $0.0035$ (LiteBIRD) and $0.0030$ (PICO), with recovered values matching true parameters within $1\sigma$  across the entire parameter space of the test set. The agreement between $\langle \sigma \rangle$ and  the $1\sigma$ ({\sc rmse}) confirms the internal consistency and stability of the method.
Moreover, our LiteBIRD results align with the latest mission forecasts \citep{LiteBIRD:2022cnt}.

The uncertainties reported here should be interpreted as baseline performance under idealized full-sky conditions. They represent lower-bound estimates in the absence of large-scale systematics such as $1/f$ noise or other effects related to the scan strategy. The primary objective of this Letter is to establish the methodological soundness and statistical consistency of the integrated pipeline, rather than to deliver a final experimental forecast including all instrumental effects. Extensions to partial-sky analysis, correlated and non-Gaussian noise, realistic foreground complexity are natural next steps.
The computational efficiency of the ABS–NN framework makes it well suited to accommodate these extensions without prohibitive cost, as the pipeline can be retrained on updated simulation ensembles as increasingly realistic conditions are incorporated. However, these additions require a consistent treatment of E–B leakage and mask-induced mode coupling that would introduce additional sources of uncertainty unrelated to the core methodological contribution presented here.


In summary, the synergy between ABS’s speed and NN inference establishes a scalable paradigm for simulation-based cosmological inference in forthcoming CMB missions, alternative to traditional likelihood analyses. By demonstrating that component separation and ML-based parameter estimation can be efficiently combined, this work opens a new practical direction for end-to-end analysis pipelines in precision cosmology.


\begin{acknowledgments}

L. S. is supported by the National Key R\&D Program of China (2020YFC2201600). C. P. N. and E. F. thank the Serrapilheira Institute for financial support. This research used computing resources
at Kavli IPMU. The Kavli IPMU is supported by the WPI (World Premier International Research Center) Initiative of the MEXT (Japanese Ministry of Education, Culture, Sports, Science and Technology).  C. B.  acknowledges partial support by the Italian Space Agency LiteBIRD Project (ASI Grants No. 2020-9-HH.0 and 2016-24-H.1-2018), and the Italian Space Agency Euclid Project, as well as the InDark and LiteBIRD Initiative of the National Institute for Nuclear Physiscs, and the RadioForegroundsPlus Project HORIZON-CL4-2023-SPACE-01, GA 101135036, and Project SPACE-IT-UP  by the Italian Space Agency and Ministry of University and Research, Contract Number  2024-5-E.0 and The CMB-Inflate project funded by the European Union’s Horizon 2020 Research and Innovation Staff Exchange under the Marie Skłodowska-Curie grant agreement No 101007633.
\end{acknowledgments}

\bibliography{refs}{}
\bibliographystyle{aasjournalv7}



\end{document}